\documentclass[aps,prl,reprint,superscriptaddress,twocolumn,longbibliography,floatfix]{revtex4-2}
\usepackage{amsfonts,amssymb,amscd,amsthm}
\usepackage{graphicx}
\usepackage{mathrsfs}
\usepackage{color}
\usepackage[intlimits]{amsmath}
\usepackage[colorlinks, citecolor=red]{hyperref}
\usepackage{etoolbox}

\theoremstyle{definition}
\newtheorem*{theorem*}{Theorem}

\begin{document}
\title{Frequency-Multiplexed Parallel Gates for Quantum LDPC Codes in a Two-Dimensional Ion Crystal}

\author{G.-X. Tang}
\affiliation{Center for Quantum Information, Institute for Interdisciplinary Information Sciences, Tsinghua University, Beijing 100084, PR China}

\author{L.-M. Duan}
\email{lmduan@tsinghua.edu.cn}
\affiliation{Center for Quantum Information, Institute for Interdisciplinary Information Sciences, Tsinghua University, Beijing 100084, PR China}
\affiliation{Hefei National Laboratory, Hefei 230088, PR China}
\affiliation{New Cornerstone Science Laboratory, Institute for Interdisciplinary Information Sciences, Tsinghua University, Beijing 100084, PR China}

\author{Y.-K. Wu}
\email{wyukai@mail.tsinghua.edu.cn}
\affiliation{Center for Quantum Information, Institute for Interdisciplinary Information Sciences, Tsinghua University, Beijing 100084, PR China}
\affiliation{Hefei National Laboratory, Hefei 230088, PR China}

\begin{abstract}
Quantum low-density parity-check (qLDPC) codes admit high encoding rates but require nonlocal entangling gates for syndrome measurement. Instead of physically moving the qubits which slows down with the increasing qubit number, here we propose to achieve parallel nonlocal entangling gates on a two-dimensional (2D) ion crystal using frequency-multiplexing. Adiabatic conditions ensure the suppression of gate infidelity and crosstalk error, as well as their robustness against slow drift in the trap frequency which is a leading error source in ion trap. We consider a numerical example of a $[[248,10,18]]$ bivariate bicycle code on a 2D crystal of 512 ions. By optimizing the mapping of the qubits and the assignment of the frequency bands for multiplexing, we show that a moderate laser power is sufficient for parallelism, and that a logical error rate of $10^{-12}$ can be achieved under realistic noise parameters.
\end{abstract}

\maketitle

\emph{Introduction.---} For a universal quantum computer to execute algorithms of arbitrary size, quantum error correction (QEC) is the indispensable cornerstone that achieves sufficiently low logical error rates from the faulty physical operations \cite{nielsen2000quantum}. Among the various quantum error correction codes, the surface code is well-known for its high threshold and its two-dimensional (2D) layout \cite{PhysRevA.86.032324} which maps naturally onto the superconducting qubit architecture \cite{Acharya2025,rqkg-dw31}. However, this planar geometry comes at the cost of a large qubit overhead, as the encoding rate vanishes with the increasing code distance \cite{bravyi2010_tradeoffs_2d_storage}. The past few years have seen a surge of interest in quantum low-density parity-check (qLDPC) codes \cite{breuckmann2021_quantum_ldpc_overview} because they promise a much higher qubit efficiency while maintaining a high error threshold \cite{tillich2014_quantum_ldpc_hypergraph_product,panteleev2022_asymptotically_good_quantum_ldpc,leverrier2022_quantum_tanner_codes,bravyi2024_high_threshold_bb_memory,wang2026_low_overhead_qec_codes,cain2026_shor_10000_reconfigurable_atoms,lu2026quantumerrorcorrectionultralow}. To achieve the required nonlocal qubit connectivity of the qLDPC codes, currently the mainstream strategy is to physically move the qubits such as neutral atoms in optical tweezer arrays \cite{Beugnon2007,bluvstein2024_logical_quantum_processor_atom_arrays} and trapped ions in quantum charge-coupled device (QCCD) architectures \cite{Kielpinski2002,ransford2026_98_qubit_helios}. Nevertheless, qubit shuttling is inherently slow and error-prone, as the time required to achieve all-to-all connectivity grows with the qubit number. This can increase the effective physical error rate per QEC cycle and degrade QEC performance.

Recently, 2D ion crystals have been proposed as an alternative way to scale up universal quantum computing \cite{guo2024_site_resolved_2d_ion_simulator,szymanski2012_large_2d_coulomb_crystals,wang2020_coherently_manipulated_2d_ion_crystal,kato2022_two_tone_doppler_radial_2d_crystals,kiesenhofer2023_controlling_2d_coulomb_crystals,schroer2026versatilelasermachinedrftrap}. With the ions staying in their equilibrium positions, distant ions can be entangled by their laser-induced coupling to the collective phonon modes. Such all-to-all entangling gates have been achieved for a small crystal of four ions as a proof-of-principle demonstration \cite{hou2024_entangling_gates_2d_ion_crystal}, as well as in 1D for up to 30 qubits \cite{Chen2024benchmarkingtrapped}. However, due to the fact that the collective phonon modes involve all the ions in a crystal, it is not clear if parallel entangling gates can be achieved on a large 2D ion crystal with low crosstalk errors. Existing experimental demonstrations at small scales design the parallel gates by a brute-force global optimization \cite{figgatt2019_parallel_entangling_operations,lu2019_global_entangling_gates}, which is NP-hard and not scalable to large crystals. More efficient numerical algorithms have later been developed to design the parallel entangling gates in polynomial time using multiple time segments \cite{grzesiak2020_efficient_simultaneously_entangling_gates} or multiple frequency tones \cite{huang2026_large_scale_multimode_gate_synthesis}. In these algorithms, however, the suppression of the two-qubit crosstalk among different gate pairs is only achieved for the ideal pulse sequence, while practical pulse imperfections or experimental noise can still lead to crosstalk errors which are detrimental for QEC \cite{liu2025_parallel_gate_crosstalk}. It is thus desirable to develop a scheme for parallel entangling gates with robustness against experimental noise, and to illustrate that sufficiently low logical error rates can be obtained under realistic experimental parameters, so as to justify the feasibility of quantum computing with large 2D ion crystals.

Here we propose a frequency-multiplexing scheme for parallel entangling gates on a large 2D ion crystal. We employ the adiabatic elimination of spin-motion entanglement (AESE) pulse design \cite{sutherland2024_laser_free_aese,hughes2025trapped,x9b9-5d78} to disentangle all the phonon modes simultaneously without the need of complicated waveform optimization, and we assign separated frequency bands to run multiple entangling gates in parallel with a robust suppression of their crosstalk errors. Noticeably, the gate time stays constant independent of the ion number or the distance between the targeted ions, as long as sufficient laser power can be applied. This makes an advantage over physically transporting the qubits as the system scales up. We further analyze a concrete example of encoding the $[[248,10,18]]$ bivariate bicycle code (BB18) \cite{rmy6-9n89,cain2026_shor_10000_reconfigurable_atoms} in a 2D crystal of 512 ions, and show that a logical error rate of $10^{-12}$ can be obtained under realistic noise parameters. By optimizing the mapping of the qubits and the assignment of the frequency bands, the required per-ion laser power is moderate and is much lower than that in the state-of-the-art optical tweezer array experiments.

\begin{figure}[!tbp]
\includegraphics[width=\columnwidth]{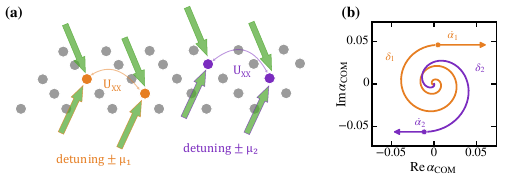}
\caption{\label{fig:scheme}
Illustration of the frequency-multiplexed parallel entangling gate. (a) Target ions are coupled by individual Raman laser beams to the drumhead phonon modes of a 2D ion crystal. All the parallel entangling gates share the same amplitude ramping waveform, and occupy different frequency bands as their laser detuning. The waveform and the frequency bands are designed to satisfy the adiabatic conditions for high gate fidelity and low crosstalk. (b) Ions involved in different entangling gates have distinct rotation speeds in the phase space. For large frequency separation and long ramping time, the crosstalk error between the two gates vanishes, which is proportional to the mixed area \cite{Schneider_2013} of the two phase space trajectories. In comparison, the two-qubit entangling phases of individual gates are proportional to the enclosed areas of individual curves and always accumulate with time.}
\end{figure}

\emph{Parallel entangling gates.---}
We consider a 2D crystal of $N$ ions with $N$ drumhead phonon modes. Suppose we want to entangle $K$ distinct pairs of ions labeled by $(i_1, j_1),\,(i_2, j_2),\,\cdots,\,(i_K,j_K)$ for a duration of $T$. As shown in Fig.~\ref{fig:scheme}, we shine focused laser beams on the target ions simultaneously with a global envelope $\gamma(t)=\sin^2(\pi t/T)$ ($0\le t\le T$) \cite{sutherland2024_laser_free_aese} to generate a spin-dependent force (SDF). Throughout this work we take the parameters of counter-propagating bichromatic $355\,$nm Raman laser beams for the manipulation of ${}^{171}\mathrm{Yb}^+$ ions as an illustrative example \cite{Chen2024benchmarkingtrapped,debnath2016programmable}, although other ion species or other gate schemes like the light-shift gate \cite{leibfried2003experimental,baldwin2021_light_shift_gate} can work equally well. The SDF Hamiltonian can be written as
\begin{equation}
\begin{aligned}
H(t) = {}&
 \gamma(t)\sum_{k=1}^K \Omega_k\sin \mu_k t\sum_{l\in\{i_k,j_k\}}
\sum_{m}\eta_m b_{lm}\sigma_l^x \\
& \quad\times\left(a_m e^{-i\omega_m t}+a_m^\dagger e^{i\omega_m t}\right)
,
\label{eq:Hamiltonian}
\end{aligned}
\end{equation}
where $\Omega_k$ is the effective Raman Rabi rate for the gate $k$ on ions $i_k$ and $j_k$, $\mu_k$ is the symmetric detuning of the bichromatic Raman laser from the carrier transition \cite{debnath2016programmable}, $\sigma_l^x$ is the Pauli $X$ operator of the $l$-th spin, $\eta_m$ is the Lamb-Dicke parameter of the $m$-th phonon mode with frequency $\omega_m$ and annihilation and creation operators $a_m$ and $a_m^\dag$, and $b_{lm}$ is the normalized mode vector of the ion $l$ in the phonon mode $m$.

Denote $\delta_{km}\equiv\mu_k-\omega_m$ as the laser detuning of the gate $k$ from the $m$-th phonon mode. For sufficiently slow amplitude ramping $|\delta_{km}|T\gg 1$, the residual spin-phonon entanglement is suppressed to the order of $O(\eta_m b_{lm}\Omega_k/(\delta_{km}^3 T^2))$ \cite{sutherland2024_laser_free_aese}. Besides, given a gate time $T$, the Rabi rate $\Omega_k$ for a maximal entangling gate is designed to set the two-qubit entangling phase $\Theta_{i_k,j_k}\propto \Omega_k^2 T\sum_m \eta_m^2 b_{i_k,m}b_{j_k,m}/\delta_{km}$ to be $\pm\pi/4$, such that the ideal unitary evolution $\exp(i\Theta_{i_k,j_k}\sigma_{i_k}^x\sigma_{j_k}^x)$ is equivalent to a CNOT gate up to single-qubit rotations. Combining them, we can estimate the gate infidelity due to the residual spin-phonon entanglement under the ideal pulse sequence to be $O(1/(\delta_{\mathrm{min}} T)^5)$ where $\delta_{\mathrm{min}}$ is the minimal laser detuning to all the phonon modes. As for the crosstalk between two pairs of parallel gates $k$ and $l$, it is characterized by the nonzero two-qubit entangling phases $\Theta_{pq}$ with $p\in\{i_k,j_k\}$ and $q\in\{i_l,j_l\}$, proportional to the mixed area \cite{Schneider_2013} of the phase space trajectories of individual gates as shown in Fig.~\ref{fig:scheme}(b). By setting a $\Delta\mu$ difference between their laser detunings, we can estimate such two-qubit phases to be $O(1/[(\delta_{\mathrm{min}} T)(\Delta\mu T)^5])$ when $\Delta\mu \lesssim \delta_{\mathrm{min}}$, and $O(1/[(\delta_{\mathrm{min}} T)^3(\Delta\mu T)^3])$ when $\Delta\mu \gtrsim \delta_{\mathrm{min}}$, under the rotating-wave approximation (RWA). The detailed formulae can be found in Supplemental Material \cite{supp}.

\begin{figure}[!tbp]
\includegraphics[width=\linewidth]{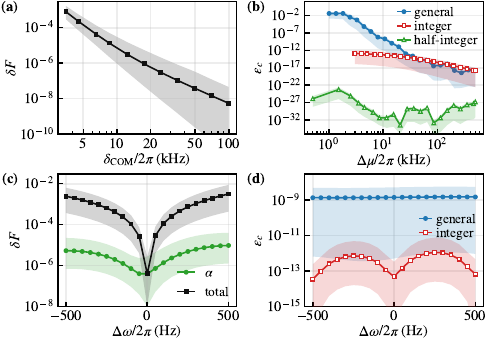}
\caption{\label{fig:gate}
Performance of the frequency-multiplexed parallel entangling gates on a 2D crystal of $N=512$ ions with a gate duration $T=1\,$ms. (a) Gate infidelity $\delta F$ vs. detuning $\delta_{\mathrm{COM}}$. (b) Crosstalk error $\epsilon_c$ between two gate pairs vs. the separation $\Delta\mu$ between their frequency bands. Blue circles represent general parameter settings, and red squares and green triangles are the sweet spots satisfying $\Delta\mu T=2\pi r$ ($r \in \mathbb{Z}, r\ge 3$) and $\Delta\mu T=2\pi (r+1/2)$ ($r \in \mathbb{Z}, r\ge 0$), respectively. (c) Total gate infidelity $\delta F$ (black squares) and that due to residual spin-phonon entanglement (green circles) vs. drift of the trap frequency $\Delta\omega$. The ideal gate parameters are designed for $\delta_{\mathrm{COM}}/2\pi=10\,$kHz. (d) Crosstalk between two gate pairs vs. trap frequency drift $\Delta\omega$. The ideal gate parameters are placed at a sweet spot $\Delta\mu/2\pi=10\,$kHz (red squares) or at general positions $\Delta\mu/2\pi=10.1\,$kHz (blue circles).
Each data point is averaged over 1000 samples of random ion pairs (for gate infidelity) or random pairs of gates (for crosstalk), and the shaded regions represent the 95\% equal-tailed interval.}
\end{figure}

As a numerical example, we consider a 2D crystal of $N=512$ ions with a trap frequency of $\omega_z/2\pi=2.164\,$MHz perpendicular to the plane (following the parameters used in Ref.~\cite{guo2024_site_resolved_2d_ion_simulator}) and fix a gate duration of $T=1\,$ms. In Fig.~\ref{fig:gate}(a) we design the required Raman Rabi rates for randomly chosen ion pairs when the laser detuning is placed above the center-of-mass (COM) mode by $\delta_{\mathrm{COM}}=\mu-\omega_z$, and calculate the corresponding gate infidelity. The black data points show the average infidelity over 1000 random ion pairs, while the shaded region represents 95\% equal-tailed interval for all the samples. Similarly, in Fig.~\ref{fig:gate}(b) we randomly choose two pairs of gates $A=\{i_A,j_A\}$ and $B=\{i_B,j_B\}$, and set their detunings to be $\delta_{\mathrm{COM}}/2\pi=10\,$kHz and $\delta_{\mathrm{COM}}+\Delta\mu$, respectively. For general frequency settings (blue circles), we obtain the $1/(\Delta\mu T)^5$ scaling in the two-qubit phase and hence $1/(\Delta\mu T)^{10}$ in the crosstalk error $\epsilon_c\equiv\sum_{p\in A,q\in B} |\Theta_{pq}|^2$. Additionally, we observe two sets of sweet spots at $\Delta\mu T=2\pi r$ ($r \in \mathbb{Z}, r\ge 3$) (red squares) and $\Delta\mu T=2\pi (r+1/2)$ ($r \in \mathbb{Z}, r\ge 0$) (green triangles): The latter have vanishing crosstalk under RWA and the remaining errors are governed by the counter-rotating terms; The former are local minima at small $\Delta\mu$, and are more preferable for parallelizing multiple gates while maintaining the frequency separation between any two gates to locate on a sweet spot. More details can be found in Supplemental Material \cite{supp}.

We further examine the robustness of the gate performance against slow drift $\Delta\omega$ in the trap frequency, which is currently a leading error source for entangling gates in ion trap \cite{PhysRevApplied.22.014007}. As shown in Fig.~\ref{fig:gate}(c), the decoupling of the spins and the phonons (green circles) is robust as long as the adiabatic condition $\delta_{\mathrm{COM}}T\gg 1$ is satisfied, while the total gate infidelity (black squares) is dominated by the change in the two-qubit entangling phase and follows a $(\Delta\omega/\delta_{\mathrm{COM}})^2$ scaling. As for the crosstalk error, the performance under general parameters (blue circles) is insensitive to the trap frequency drift when $\Delta\omega\ll\Delta\mu$. In contrast, the ``integer'' sweet spots (red squares) do see the effect of the drift, but the crosstalk error $\epsilon_c$ remains negligible (below $10^{-11}$) under typical gate parameters.

\begin{figure}[!tbp]
\includegraphics[width=\columnwidth]{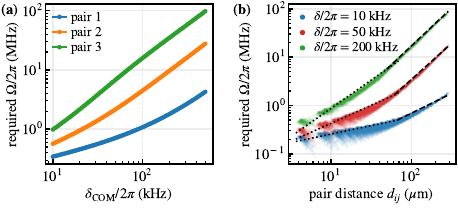}
\caption{\label{fig:intensity}
Scaling of required laser intensity on a 2D crystal of $N=512$ ions with a gate duration $T=1\,$ms. (a) Required Raman Rabi rate $\Omega$ vs. detuning $\delta_{\mathrm{COM}}$ for three representative ion pairs at the distance of $d=19.3,68.9,157.6\,\mu$m. (The average distance between nearest-neighbor ions is $4.3\,\mu$m.) (b) Required Raman Rabi rate $\Omega$ vs. ion distance $d$ under three representative laser detunings $\delta_{\mathrm{COM}}/2\pi=10,50,200\,$kHz. Dotted and dashed lines show power-law fits (linear on the log-log plot) for the small-$d$ and large-$d$ regimes, respectively.}
\end{figure}

\emph{Scaling of laser intensity.---}
The above analysis suggests that a larger detuning $\delta_{\mathrm{COM}}$ and a larger frequency band separation $\Delta\mu$ are desirable for higher gate fidelity, lower crosstalk error and better robustness against parameter drift. However, the cost to pay is a higher laser intensity under a given gate time, which can become a limitation when a high parallelism level is targeted. Fig.~\ref{fig:intensity}(a) shows this scaling of the required Raman Rabi rate $\Omega$ versus the detuning $\delta_{\mathrm{COM}}$ for three representative ion pairs. At small detuning and for nearby ions, the two-qubit entangling phase mainly comes from a few phonon modes close to the COM mode, so that a $\Omega\propto\sqrt{\delta_{\mathrm{COM}}}$ scaling is observed. At large detuning comparable to the phonon bandwidth and for distant ions, a dipole-dipole interaction appears through the cancellation between different phonon modes, and we obtain the scaling $\Omega\propto d^{3/2}(\mu^2-\omega_z^2)=d^{3/2}\delta_{\mathrm{COM}}(2\omega_z+\delta_{\mathrm{COM}})$ (see Supplemental Material for details \cite{supp}).

In Fig.~\ref{fig:intensity}(b) we further plot the scaling of $\Omega$ versus the distance $d$ between ion pairs under three representative laser detunings. Again for small detuning and small ion distance the interaction is dominated by a few modes close to the COM mode, so that the required laser intensity shows only weak spatial dependence (a slope of $0.35$ for the small-$d$ regime when $\delta_{\mathrm{COM}}/2\pi=10\,$kHz). For large detuning and large ion distance, again the $\Omega\propto d^{3/2}$ scaling of the dipole-dipole interaction appears, as shown by the large-$d$ regime when $\delta_{\mathrm{COM}}/2\pi=50,\,200\,$kHz.

From Fig.~\ref{fig:intensity} it seems that a large ion distance with a simultaneous large laser detuning requires a Raman Rabi rate of up to $10^2\,$MHz, which is beyond the current experimental value and may lead to the breakdown of the Lamb-Dicke approximation. However, this can be avoided by a suitable arrangement of the frequency bands such that distant entangling gates always occupy the lowest bands with small detunings while adjacent ions take higher ones. In the following, we demonstrate this idea by compiling the syndrome measurement circuit of a QEC code encoded in a 2D ion crystal.

\emph{Demonstration with the BB18 qLDPC code.---}
We consider the $[[248,10,18]]$ bivariate bicycle code (BB18) \cite{rmy6-9n89,cain2026_shor_10000_reconfigurable_atoms} which requires 248 data qubits and 248 measurement qubits (124 for $X$-type stabilizers and 124 for $Z$-type stabilizers which purposely include redundancy \cite{bravyi2024_high_threshold_bb_memory}). For simplicity, we assume ideal single-qubit rotations and focus on the scheduling of two-qubit gates which dominate the gate errors. Since each stabilizer has a weight of 6, we need totally 1488 two-qubit entangling gates in each round of syndrome measurement, which can be divided into 12 groups of equal size that act on disjoint ion pairs \cite{cain2026_shor_10000_reconfigurable_atoms}. Following Ref.~\cite{cain2026_shor_10000_reconfigurable_atoms}, we consider logical $X$ operators by initializing all the data qubits in $|+\rangle$, and perform $d/2=9$ rounds of syndrome measurements. To decode the corresponding logical error rate $p_L$ that any of the 10 logical $X$ operators are decoded incorrectly, only the $X$-type measurement results are needed, although during each QEC cycle we still measure all the $Z$-type and $X$-type stabilizers.

We map the required 496 physical qubits to the 512-ion crystal with 16 ions unused which may provide sympathetic cooling together with the measurement qubits. The mid-circuit detection of the measurement qubits can be achieved without affecting the data qubits using the dual-type scheme \cite{yang2022realizing,10.1063/5.0069544}, where the data qubits can be temporarily converted to $F_{7/2}$ hyperfine levels while the remaining qubits in $S_{1/2}$ can be measured with high fidelity by electron shelving. The mapping $\pi:\{q_1,\,\cdots,\,q_{496}\}\to \{1,\,\cdots,\,512\}$ from the qubits to the ions in the crystal is chosen to minimize the total gate infidelity $C(\pi)\equiv\sum_{(u,v)\in E} \delta F(\pi(u),\pi(v))$ where $E$ is the set of two-qubit pairs on which entangling gates need to be performed during the syndrome measurement, and $\delta F$ is the gate infidelity for an ion pair under the same laser detuning and the same level of trap frequency drift as shown in Fig.~\ref{fig:gate}(c). Note that this optimization also roughly corresponds to mapping more entangling gates onto nearby ion pairs as they require lower laser intensity and hence generate weaker phonon excitations. This is a quadratic assignment problem \cite{koopmans1957_assignment_location}, and we solve it approximately using the fast approximate quadratic assignment algorithm \cite{vogelstein2015_fast_approximate_qap}.

After fixing the qubit mapping, we next consider the scheduling of the parallel entangling gates. As described above, the required two-qubit gates for one QEC round can be divided into 12 groups, each containing 124 gates on disjoint ions which can be performed in parallel. We therefore define 124 frequency bands in our frequency-multiplexed parallel entangling gate scheme. As a balance between noise robustness and laser intensity, we set the lowest band at $\delta_{\mathrm{COM}}/2\pi=10\,$kHz above the COM mode, and choose the separation between adjacent frequency bands to be $\Delta\mu/2\pi=3.5\,$kHz. For this setting to locate on a sweet spot, we use a gate time $T=6\pi/\Delta\mu \approx 857\,\mu$s. Once we assign an ion pair to a frequency band, its Raman Rabi rate is determined by requiring the two-qubit entangling phase to be $\pm\pi/4$ after the given gate duration. Then we can optimize the required laser intensity by suitably assigning parallel gates into different frequency bands as mentioned before. Further saving in the laser power can be achieved by a reduction in the parallelism level, which uses less frequency bands but increases the total duration of a QEC cycle and hence enlarges the idling error \cite{liu2025_parallel_gate_crosstalk}. We divide the aforementioned 12 groups of disjoint gates into $S\ge 12$ time slots, and optimize the assignment $\kappa$ of the time and the frequency slots so as to minimize the peak laser intensity over different layers $P(\kappa)\equiv \max_{1\le t\le S} \sum_i \Omega(\kappa;t,i)$ where $\Omega(\kappa;t,i)$ is the required Raman Rabi rate of the $i$-th gate at the $t$-th time slot under a given assignment $\kappa$. The detailed optimization algorithm can be found in Supplemental Material \cite{supp}.

\begin{figure}[!tbp]
\includegraphics[width=\columnwidth]{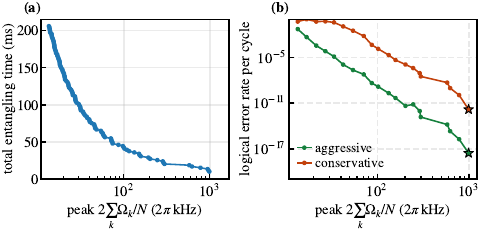}
\caption{\label{fig:QEC}
Performance of BB18 code on a 2D crystal of $N=512$ ions. (a) Total entangling gate time per syndrome-measurement cycle vs. peak per-ion Rabi rate $2\sum_k\Omega_k/N$ for different scheduling of the parallel gates. (b) Logical error rate per syndrome-measurement cycle vs. peak per-ion Rabi rate $2\sum_k\Omega_k/N$. We consider two representative sets of noise parameters. A more conservative set (red circles) assumes a trap frequency drift $\Delta\omega/2\pi=\pm 500\,$Hz, a spin dephasing time $T_2=1\,$s, an average phonon number $\bar{n}_m=1$ for each drumhead phonon mode, and a measurement error of $p_{\mathrm{meas}}=10^{-3}$. A more aggressive set of parameters (green circles) uses $\Delta\omega/2\pi=\pm 100\,$Hz, $T_2=3\,$s, $\bar{n}_m=0.1$ and $p_{\mathrm{meas}}=5\times 10^{-4}$.}
\end{figure}

In Fig.~\ref{fig:QEC}(a), we present the competition between the QEC cycle duration and the required laser power by tuning the number of time slots $S$ and hence the maximal parallelism level $q$. For the highest parallelism level $q=124$ we need a peak per-ion Raman Rabi rate $2\sum_k\Omega_k/N=2\pi\times0.99\,$MHz, where the index $k$ runs over all the parallel gate pairs at the most demanding time slot, and the factor of $2$ accounts for the two ions involved in one entangling gate. This value is much lower than the typical trap depth in current neutral atom experiments using optical tweezer arrays \cite{Bluvstein2022,Graham2022}. Note that the Rabi rates are not uniform among different ion pairs, and we find the highest value to be $\Omega=2\pi\times4.27\,\mathrm{MHz}$. Such a Raman Rabi rate has been achieved for the ${}^{43}\mathrm{Ca}^+$ ions while still maintaining a high gate fidelity \cite{Schafer2018}. Also note that the total duration of the $S=12$ layers of entangling gates is about $10.3\,$ms, significantly longer than the $1$-$2\,$ms qubit state readout which has been demonstrated for hundreds of ions \cite{guo2024_site_resolved_2d_ion_simulator}. Therefore in the following we only use the total two-qubit gate duration to estimate the idling error in the QEC circuit.

To examine the logical performance of the BB18 code on the 2D ion crystal, we simulate the QEC circuit by a Clifford circuit simulator and estimate the logical error rate $p_L$ under a Relay-BP decoder \cite{muller2025_relay_bp_realtime_memory}. Our error model includes the two-qubit gate error and the crosstalk error due to a given level of trap frequency drift $\Delta\omega$ and a given average phonon number $\bar{n}$ \cite{liu2025_parallel_gate_crosstalk}, the idling error on all the qubits due to a given $T_2$ coherence time, and the measurement error $p_{\mathrm{meas}}$ (see Supplemental Material for details).
The numerical simulation results are shown in Fig.~\ref{fig:QEC}(b) corresponding to the scheduling of the parallel gates in Fig.~\ref{fig:QEC}(a). We consider two sets of noise parameters, a more conservative one (red) and a more aggressive one (green), both of which have been realized at small scales (see, e.g. Refs. \cite{10.1063/1.4948734,Egan2021,PhysRevLett.129.140501,PhysRevA.81.040302}). When the logical error rates are below $10^{-5}$, we choose to scale up all the physical error rates uniformly by a factor $\lambda>1$, fit $\log p_L(\lambda)=a+b\log\lambda$, and extrapolate to the desired point $\lambda=1$ \cite{bonillaataides2021_xzzx_surface_code,pecorari2025_high_rate_ldpc_neutral_atoms,cain2026_shor_10000_reconfigurable_atoms}. Because the crosstalk error is largely suppressed at the sweet spot and becomes subdominant, the effect of reducing the required laser intensity by limiting the parallelism level is mainly to enlarge the idling error and hence to increase the logical error rate. To achieve a logical error rate of, say, $10^{-12}$, we estimate a peak per-ion Raman Rabi rate of $2\pi\times0.99\,$MHz and $2\pi\times0.29\,$MHz for the conservative and the aggressive error parameters, respectively.

\emph{Discussion.---}
To sum up, in this work we propose a frequency-multiplexed scheme for parallel entangling gates on a 2D ion crystal. We utilize adiabatic pulse design to suppress the infidelity of individual entangling gates and the crosstalk error between parallel gates which occupy distinct frequency bands, and to achieve noise-robustness against slow trap frequency drift. For our numerical example with $N=512$ ions, we show that a moderate laser intensity, much lower than that in the state-of-the-art optical tweezer array experiments \cite{Bluvstein2022,Graham2022}, is sufficient for quantum error correction with the $[[248,10,18]]$ bivariate bicycle code to achieve a logical error rate of $10^{-12}$, so as to support practical applications under realistic noise parameters.

\emph{Data availability.---}
The data that support the findings of this article are openly available in Zenodo \cite{tang2026_data}. 

\begin{acknowledgments}
This work was supported by the National Natural Science Foundation of China (Grant No. 12575021), the Quantum Science and Technology-National Science and Technology Major Project (2021ZD0301601), the Tsinghua University Initiative Scientific Research Program, and the Ministry of Education of China. L.-M. D. acknowledges in addition support from the New Cornerstone Science Foundation through the New Cornerstone Investigator Program. Y.-K. W. acknowledges in addition support from Tsinghua University Dushi program.
\end{acknowledgments}

\end{document}


\makeatletter
\renewcommand{\thefigure}{S\arabic{figure}}
\renewcommand{\thetable}{S\arabic{table}}
\renewcommand{\theequation}{S\arabic{equation}}
\makeatother

\title{Supplemental Material for
``Frequency-Multiplexed Parallel Gates for Quantum LDPC Codes in a Two-Dimensional Ion Crystal''}

\author{G.-X. Tang}
\affiliation{Center for Quantum Information, Institute for Interdisciplinary Information Sciences, Tsinghua University, Beijing 100084, PR China}

\author{L.-M. Duan}
\email{lmduan@tsinghua.edu.cn}
\affiliation{Center for Quantum Information, Institute for Interdisciplinary Information Sciences, Tsinghua University, Beijing 100084, PR China}
\affiliation{Hefei National Laboratory, Hefei 230088, PR China}
\affiliation{New Cornerstone Science Laboratory, Institute for Interdisciplinary Information Sciences, Tsinghua University, Beijing 100084, PR China}

\author{Y.-K. Wu}
\email{wyukai@mail.tsinghua.edu.cn}
\affiliation{Center for Quantum Information, Institute for Interdisciplinary Information Sciences, Tsinghua University, Beijing 100084, PR China}
\affiliation{Hefei National Laboratory, Hefei 230088, PR China}

\maketitle

\section{Analytical expressions for crosstalk error between frequency-multiplexed AESE gates}
Consider two gates $A$ and $B$ executed in parallel:
\begin{equation}
\begin{aligned}
H(t) ={}&
\Omega_A\gamma(t)\sin(\mu_A t)
\sum_{j\in A}\sum_m\eta_m b_{jm}X_j
\left(a_m e^{-i\omega_m t}+a_m^\dagger e^{i\omega_m t}\right)
\\
&+
\Omega_B\gamma(t)\sin(\mu_B t)
\sum_{j\in B}\sum_m\eta_m b_{jm}X_j
\left(a_m e^{-i\omega_m t}+a_m^\dagger e^{i\omega_m t}\right).
\end{aligned}
\label{eq:aese_general_hamiltonian}
\end{equation}

Here, the sums over \(j\in A\) and \(j\in B\) run over the ions on which gates \(A\) and \(B\) act, respectively. \(\Omega_A\) and \(\Omega_B\) are the peak effective Raman Rabi rates, \(\gamma(t)\) is their common dimensionless envelope, \(b_{jm}\) is the normalized mode vector of ion \(j\) in phonon mode \(m\), and \(\eta_m=|\Delta k|/\sqrt{2M\omega_m}\) is the Lamb--Dicke factor associated with the wave-vector difference \(\Delta k\)~\cite{wu2018_noise_analysis_raman_gates}. Residual spin--motion entanglement for each individual drive is adiabatically suppressed by smooth switching, as described in Ref.~\cite{sutherland2024_laser_free_aese}. Here we focus on the two-spin phase generated by the interaction between the two drives. Because the Hamiltonian is linear in the bosonic operators, the Magnus expansion terminates at second order. Using
\begin{equation}
\left[
a_m e^{-i\omega_m t}+a_m^\dagger e^{i\omega_m t},
a_m e^{-i\omega_m t'}+a_m^\dagger e^{i\omega_m t'}
\right]
=-2i\sin[\omega_m(t-t')],
\label{eq:aese_boson_commutator}
\end{equation}
the crosstalk $e^{i\Theta X_r X_s}$ accumulated between a qubit \(r\in A\) and a qubit \(s\in B\) is
\begin{equation}
\resizebox{0.96\textwidth}{!}{$\displaystyle
\Theta_{rs}^{AB}
=\Omega_A\Omega_B
\sum_m\eta_m^2b_{rm}b_{sm}
\int_0^Tdt\int_0^t dt'\,
\gamma(t)\gamma(t')
\sin[\omega_m(t-t')]
\left[
\sin(\mu_A t)\sin(\mu_Bt')
+\sin(\mu_B t)\sin(\mu_A t')
\right].
$}
\label{eq:aese_xtalk_exact}
\end{equation}
We now specialize to the envelope \(\gamma(t)=\sin^2(\pi t/T)\) and use the rotating-wave approximation (RWA) to retain the terms oscillating at the drive--mode detunings. For a given mode, define the dimensionless detuning and band separation by $a_m=(\mu_A-\omega_m)T$ and $d=(\mu_B-\mu_A)T$. The crosstalk phase within the RWA is then
\begin{equation}
\Theta_{rs}^{AB,\mathrm{RWA}}
=\Omega_A\Omega_BT^2
\sum_m\eta_m^2b_{rm}b_{sm}I(a_m,d),
\label{eq:aese_sin2_rwa_phase}
\end{equation}
where
\begin{equation}
\begin{aligned}
I(a,d)={}&
\frac{\pi^4}
{a(a+d)(a^2-4\pi^2)[(a+d)^2-4\pi^2]}
\\
&\times\left\{
2\cos\frac d2\,\sin\left(a+\frac d2\right)-
\frac{(2a+d)\sin d}{d}
\left[
1+\frac{2a(a+d)[3a(a+d)-d^2-20\pi^2]}
{(d^2-4\pi^2)(d^2-16\pi^2)}
\right]
\right\}.
\end{aligned}
\label{eq:aese_sin2_exact_I}
\end{equation}

For generic conditions, the trigonometric factors are assumed to be \(\mathcal{O}(1)\).
We assume that the drive frequencies are above the COM mode (and hence all the transverse modes), so that $a>0$, and label the drives such that $d>0$. The adiabatic condition guarantees \(a\gg1\). We analyze the scaling of \(I(a,d)\) in two parameter regimes. First, in the interval \(4\pi\ll d\ll a\), the denominator outside the braces approaches \(a^6\), while the coefficient multiplying \(\sin d\) inside the braces approaches \(12a^5/d^5\). The other trigonometric term inside the braces remains bounded by two, so its contribution is smaller by a factor of order \((d/a)^5\) away from zeros of \(\sin d\). Thus
\begin{equation}
I(a,d)\simeq-\frac{12\pi^4\sin d}{a\,d^5},
\qquad 4\pi\ll d\ll a.
\label{eq:aese_sin2_d5}
\end{equation}
This establishes a \(d^{-5}\) envelope in the $d \ll a$ regime. Next, take \(d\gg a\). The denominator outside the braces now approaches \(a^3d^3\), and the coefficient multiplying \(\sin d\) inside the braces approaches one. Therefore
\begin{equation}
I(a,d)\simeq
\frac{\pi^4
\left[2\cos(d/2)\sin(a+d/2)-\sin d\right]}
 {a^3d^3},
 \qquad d\gg a.
\label{eq:aese_sin2_d3}
\end{equation}
For generic \(a\), this expression gives the \(d^{-3}\) envelope. The comparison of Eqs.~(\ref{eq:aese_sin2_d5}) and (\ref{eq:aese_sin2_d3}) places the crossover between the two regimes at \(d\) of order \(a\), with the detailed oscillations determined by the exact expression.
In the numerical example shown in Fig.~2(b), \(\delta_{\rm COM}/2\pi=10\,\mathrm{kHz}\), and the \(d^{-5}\) scaling persists up to \(\Delta\mu/2\pi=50\,\mathrm{kHz}\). In the schedule shown in Fig.~4, the COM detuning \(\delta_{\rm COM}/2\pi\) spans \(10\,\mathrm{kHz}\) to \(440.5\,\mathrm{kHz}\), so \(a\) and \(d\) can be comparable. Pairs with small band separations, which are the most dangerous for crosstalk, are therefore generally in the \(d^{-5}\) regime.

In the preceding scaling analysis, we treated all trigonometric factors as \(\mathcal{O}(1)\). Their zeros can further suppress crosstalk. For example, at band separations of integer multiples of $2\pi$, the \(\sin d\) term vanishes, and the exact kernel reduces to
\begin{equation}
I(a,d)=
\frac{2\pi^4\sin a}
{a(a+d)(a^2-4\pi^2)[(a+d)^2-4\pi^2]},
\qquad
\frac{d}{2\pi}\in\mathbb Z,\quad d\ge6\pi.
\label{eq:aese_sin2_integer}
\end{equation}
The restriction excludes \(d=0,2\pi,4\pi\), which coincide with singularities of Eq.~\ref{eq:aese_sin2_exact_I}. For \(d\ll a\), Eq.~(\ref{eq:aese_sin2_integer}) is approximately \(2\pi^4\sin a/a^6\). Thus, at these integer sweet spots, \(I(a,d)\) is also \(\mathcal{O}(a^{-6})\), although it does not decay rapidly with \(d\). For \(d\gg a\gg2\pi\), it is approximately \(2\pi^4\sin a/(a^3d^3)\).

At band separations of half-integer multiples of $2\pi$, both \(\sin d\) and \(\cos(d/2)\) vanish, giving
\begin{equation}
I(a,d)=0,
\qquad \frac{d}{2\pi}\in\mathbb Z+\frac12,
\label{eq:aese_sin2_half_integer}
\end{equation}
for every \(a\). Thus the half-integer condition cancels the RWA crosstalk phase mode by mode. For convenience, in the main text we choose the smallest integer sweet spot $d=6\pi$ between adjacent frequency bands, such that the separation between any two bands also locate on an integer sweet spot. In principle, we can also choose the separation between neighboring frequency bands to be on the half-integer sweet spot $d=3\pi$. Then for more distant bands the separations will either locate on the integer or half-integer sweet spots.

\begin{figure}[h]
\centering
\includegraphics[width=\textwidth]{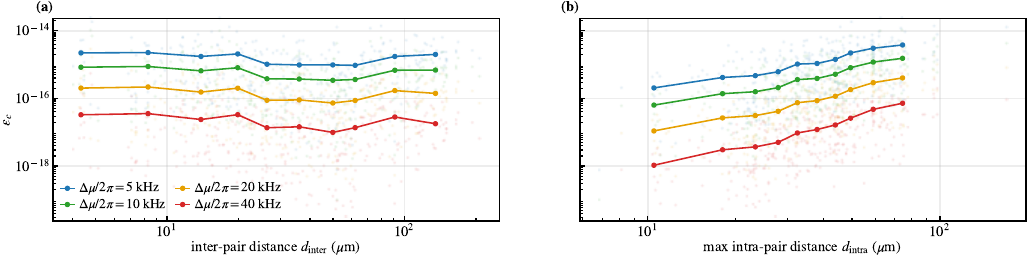}
\caption{\label{fig:raman_ms_xtalk_distance} Analytical crosstalk error between 300 randomly sampled ion pairs on the 512-ion crystal. Each faint point is one simultaneous gate-pair sample, with the crosstalk error defined as \(\epsilon_c=\sum_{p\in A,q\in B}\lvert\Theta_{pq}\rvert^2\). Solid curves are arithmetic means of \(\epsilon_c\) in distance bins. (a) Crosstalk error versus the nearest inter-pair target-ion distance \(d_{\mathrm{inter}}=\min_{i\in A,j\in B}\lVert\mathbf r_i-\mathbf r_j\rVert\), shows no systematic distance dependence. (b) Crosstalk error versus the larger intra-pair separation \(d_{\mathrm{intra}}=\max\{\lVert\mathbf r_{a_1}-\mathbf r_{a_2}\rVert,\lVert\mathbf r_{b_1}-\mathbf r_{b_2}\rVert\}\) for gate pairs \(A=(a_1,a_2)\) and \(B=(b_1,b_2)\), shows a weak positive trend.}
\end{figure}

Figure~\ref{fig:raman_ms_xtalk_distance} checks whether the pair-level crosstalk error has an additional systematic dependence on the geometry of two simultaneous gate pairs. The sampled pair-level errors show no systematic dependence on the inter-pair distance \(d_{\mathrm{inter}}\). They exhibit a weak positive dependence on the larger intra-pair distance \(d_{\mathrm{intra}}\), because more widely separated target pairs require larger calibrated effective Raman Rabi frequencies and therefore generate larger aggregate crosstalk errors. The overall crosstalk scale remains primarily controlled by the frequency-band separation \(\Delta\mu\), consistent with the kernel in Eq.~(\ref{eq:aese_sin2_rwa_phase}).

\section{Large-detuning and large-distance scaling of required laser intensity}
We now derive the scaling used in Fig.~3 of the main text. A single drive in Eq.~\ref{eq:aese_general_hamiltonian} can accumulate a phase between its two ions and generate residual spin--phonon entanglement. Its Magnus expansion gives the evolution
\begin{equation}
U(T)
=
\exp\!\left[
\sum_{l\in\{i,j\}}\sum_m X_l
\left(\alpha_{lm}a_m^\dagger-\alpha_{lm}^*a_m\right)
+i\Theta_{ij}X_iX_j
\right].
\label{eq:raman_magnus_evolution}
\end{equation}
When the envelope \(\gamma(t)\) satisfies the adiabatic conditions, the residual displacement is suppressed as \(\lvert\alpha_{lm}\rvert=O[\eta_m\lvert b_{lm}\rvert\Omega/(\lvert\delta_m\rvert^3T^2)]\), where \(\delta_m=\mu-\omega_m\)~\cite{sutherland2024_laser_free_aese,hughes2025trapped}, while the entangling phase is

\begin{equation}
\begin{aligned}
\Theta_{ij}
={}&
2\sum_m\eta_m^2b_{im}b_{jm}
\int_0^Tdt\int_0^t dt'\,
\Omega^2\gamma(t)\gamma(t')
\sin(\mu t)\sin(\mu t')
\sin[\omega_m(t-t')]
\\
\approx{}&
\Omega^2\langle\gamma^2\rangle T
\sum_m
\frac{\omega_m\eta_m^2 b_{im}b_{jm}}
{\omega_m^2-\mu^2}.
\end{aligned}
\label{eq:raman_theta}
\end{equation}
where \(\langle\gamma^2\rangle=T^{-1}\int_0^Tdt\,\gamma^2(t)\), and the second line follows from the adiabatic approximation. In this work, we determine \(\Omega\) using the full finite-time smooth-envelope integral in the first line of Eq.~(\ref{eq:raman_theta}) to avoid two-qubit-phase errors. For the scaling analysis below, however, the adiabatic expression in the second line is sufficient.

The modes considered here are transverse modes perpendicular to the plane of the two-dimensional ion crystal.  For these modes, the Raman Lamb--Dicke factor satisfies $\omega_m\eta_m^2=\frac{|\Delta k|^2}{2M}\equiv\kappa$, which is independent of the mode index.  We therefore write the spatial and spectral kernel in Eq.~(\ref{eq:raman_theta}) as
\begin{equation}
K_{ij}(\mu)
=\kappa\sum_m\frac{b_{im}b_{jm}}{\omega_m^2-\mu^2}.
\label{eq:scaling_kernel}
\end{equation}
At fixed gate duration and target angle \(\Theta_{\rm tar}\), the required peak effective Raman Rabi frequency is
\begin{equation}
\Omega_{ij}^{\rm req}
=\sqrt{
\frac{|\Theta_{\rm tar}|}
{\langle\gamma^2\rangle T|K_{ij}(\mu)|}
}.
\label{eq:scaling_required_rabi}
\end{equation}
The scaling figure uses \(|\Theta_{\rm tar}|=\pi/4\).

Let \(B\) be the orthogonal normal-mode matrix and define the mass-normalized transverse dynamical matrix
\begin{equation}
D=B\,{\rm diag}(\omega_m^2)B^{\mathsf T},
\qquad
D_{ij}=\sum_m b_{im}\omega_m^2b_{jm}.
\label{eq:scaling_dynamical_matrix}
\end{equation}
The mode sum is then the off-diagonal element of a resolvent,
\begin{equation}
\frac{K_{ij}(\mu)}{\kappa}
=\left(D-\mu^2I\right)^{-1}_{ij}.
\label{eq:scaling_resolvent}
\end{equation}
Taking \(\omega_{\rm COM}\) to be the COM-mode frequency, we introduce $L=\omega_{\rm COM}^2I-D$ and $\Lambda=\mu^2-\omega_{\rm COM}^2$. The inverse-power expansion of the resolvent is
\begin{equation}
\begin{aligned}
\left(D-\mu^2I\right)^{-1}_{ij}
&=\left[-\frac{I}{\Lambda}
+\frac{L}{\Lambda^2}
-\frac{L^2}{\Lambda^3}
+O(\Lambda^{-4})\right]_{ij}\\
&=-\frac{D_{ij}}{\Lambda^2}
-\frac{(L^2)_{ij}}{\Lambda^3}
+O(\Lambda^{-4}),
\qquad i\ne j.
\end{aligned}
\label{eq:scaling_offdiagonal_expansion}
\end{equation}
Here the large-\(\Lambda\) regime means that \(\Lambda\) suppresses the internal structure of the transverse phonon spectrum sufficiently for the first nonzero term in the target off-diagonal matrix element to control the overall behavior. The relative size of the displayed correction is characterized by \(\lvert(L^2)_{ij}/(\Lambda D_{ij})\rvert\), which can already be small for the relevant pairs while global spectral features remain visible in other matrix elements.  In this regime,
\begin{equation}
K_{ij}(\mu)
\simeq-\kappa\frac{D_{ij}}{\Lambda^2},
\qquad
\Omega_{ij}^{\rm req}
\simeq
\Lambda
\sqrt{
\frac{|\Theta_{\rm tar}|}
{\langle\gamma^2\rangle T\kappa|D_{ij}|}
}.
\label{eq:scaling_large_lambda_rabi}
\end{equation}
Thus different ion pairs share a linear high-\(\Lambda\) dependence, while \(|D_{ij}|^{-1/2}\) sets their relative offsets.

For transverse motion perpendicular to the crystal plane, the Coulomb interaction between ions \(i\) and \(j\) expands as
\begin{equation}
\frac{e^2}{4\pi\epsilon_0\sqrt{d_{ij}^2+(z_i-z_j)^2}}
=\frac{e^2}{4\pi\epsilon_0}
\left[
\frac{1}{d_{ij}}
-\frac{(z_i-z_j)^2}{2d_{ij}^3}
+O(z^4)
\right].
\label{eq:scaling_coulomb_expansion}
\end{equation}
The off-diagonal element of the mass-normalized transverse Hessian therefore obeys
\begin{equation}
|D_{ij}|=\frac{e^2}{4\pi\epsilon_0M d_{ij}^3}.
\label{eq:scaling_dipolar_dij}
\end{equation}
Combining Eqs.~(\ref{eq:scaling_large_lambda_rabi}) and (\ref{eq:scaling_dipolar_dij}) yields
\begin{equation}
\Omega_{ij}^{\rm req}
\simeq
\Lambda d_{ij}^{3/2}
\sqrt{
\frac{4\pi\epsilon_0M|\Theta_{\rm tar}|}
{\langle\gamma^2\rangle T\kappa e^2}
}
\propto\frac{\Lambda d_{ij}^{3/2}}{\sqrt{T}}.
\label{eq:scaling_combined}
\end{equation}
Note that in practical applications like the numerical example considered in the main text, we do not need to enter this parameter regime, and can achieve much lower laser intensity by optimizing the scheduling of the parallel gates.

\section{Schedule optimization of BB18 syndrome measurement circuit}
The mapping used in the main text is computed with the \texttt{quadratic\_assignment} function in SciPy's \texttt{optimize} module using \texttt{method="faq"}, which implements the fast approximate quadratic-assignment algorithm~\cite{vogelstein2015_fast_approximate_qap}.  The \(X\)- and \(Z\)-check matrices jointly define a 496-node Tanner graph, whose nodes comprise the 248 data qubits and the 124 check qubits of each type.  An edge connects a check node to a data node whenever the corresponding check matrix has a nonzero entry, and is implemented using a 2-qubit gate.  For the QAP, we pad the corresponding Tanner graph with 16 isolated nodes to match the 512-ion cost matrix and use the permutation returned by this routine.

Once the mapping is fixed, each mapped edge is assigned to a time slot and a beat-note frequency band. We schedule the \(X\)- and \(Z\)-check edges separately and concatenate the resulting schedules. The available beat-note bands are
\begin{equation}
\mu_k=\omega_{\rm COM}+\delta_{\min}+k\Delta\mu,
\qquad k=0,1,\ldots,123,
\end{equation}
and their common duration \(T\) is fixed by the integer sweet-spot condition \(\Delta\mu T=6\pi\).  For each mapped edge \(e=(i,j)\) and band \(k\), the full finite-time smooth-envelope integral is used to calibrate \(\Omega_{e,k}\) from \(|\Theta_{ij}(\Omega_{e,k},\mu_k,T)|=\pi/4\), and the scheduler uses \(\Omega_{e,k}\) directly as the assignment cost.

For the BB18 \([[248,10,18]]\) code considered in this work, each of the $X$- and $Z$-type Tanner subgraphs is bipartite and has maximum degree six.  Repeated perfect matching decomposes its edges into six conflict-free layers.
Because of the laser intensity constraints, each layer is further divided into \(r\) slots, where gates are executed simultaneously. The value of \(r\) is optimized using the method described later.

The edges of a layer are ordered by the hardness $\Omega_{e,0}$ and distributed round-robin among $r$ slots. Its gates are assigned to distinct frequency bands by solving the linear-sum assignment problem with SciPy's \texttt{linear\_sum\_assignment} function~\cite{crouse2016_rectangular_assignment}.
Specifically, let \(\phi:g\hookrightarrow\{0,1,\ldots,123\}\) denote an injective mapping from the edges in conflict-free group \(g\) to the available frequency bands. We define the group cost as
\begin{equation}
p_g=\min_{\phi:g\hookrightarrow\{0,1,\ldots,123\}}
\sum_{e\in g}\Omega_{e,\phi(e)}.
\label{eq:scheduler_linear_assignment}
\end{equation}

Dynamic programming then allocates a prescribed slot budget among the six layers.  If $C_l(r)$ is the largest group cost obtained by splitting layer $l$ into $r$ slots, the recurrence
\begin{equation}
F_l(b)=\min_r\max\{F_{l-1}(b-r),C_l(r)\},
\qquad F_0(0)=0,
\label{eq:scheduler_dynamic_program}
\end{equation}
gives the best peak cost among the constructed splits using $b$ slots.  Finally, for a total budget $S$, we choose
\begin{equation}
P_{\rm peak}(S,T)=\min_{S_X+S_Z=S}\max\{F_X(S_X),F_Z(S_Z)\}.
\label{eq:scheduler_xz_allocation}
\end{equation}
Every resulting slot is free of qubit conflicts and frequency-band collisions, and every 2-qubit gate is scheduled exactly once.  The linear-sum assignments and the dynamic program are optimal for the fixed candidate groups, whereas the layer splitting is heuristic. The scheduler thus gives a tradeoff between cycle time $TS$ and the peak Rabi frequency.

\begin{figure}[t]
\centering
\includegraphics[width=\textwidth]{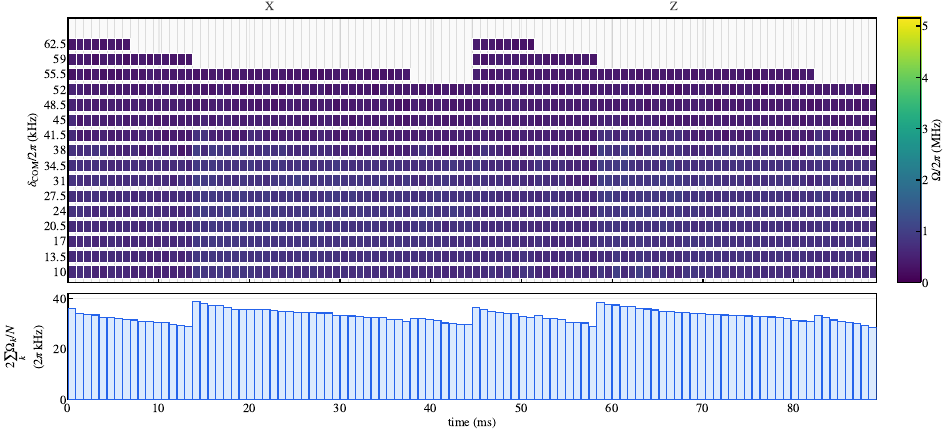}
\caption{\label{fig:raman_ms_schedule_104}
Representative schedule with \(S=104\) time slots and total entangling time \(T_{\rm cycle}=89.1\,\mathrm{ms}\).  The peak per-ion effective-Rabi allocation is \(2\sum_k\Omega_k/N=2\pi\times38.8\,\mathrm{kHz}\).  The \(X\)- and \(Z\)-check portions are labeled above the frequency-band schedule. Bar color denotes the individual gate requirement \(\Omega_k/2\pi\).}
\end{figure}

\begin{figure}[t]
\centering
\includegraphics[width=\textwidth]{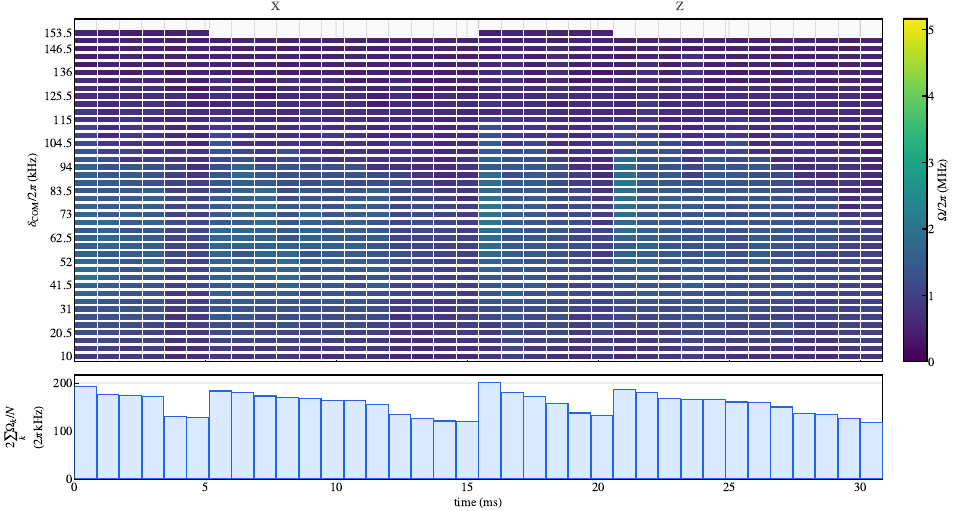}
\caption{\label{fig:raman_ms_schedule_036}
Representative schedule with \(S=36\) time slots and \(T_{\rm cycle}=30.9\,\mathrm{ms}\).  The peak per-ion effective-Rabi allocation is \(2\sum_k\Omega_k/N=2\pi\times200.9\,\mathrm{kHz}\).}
\end{figure}

\begin{figure}[t]
\centering
\includegraphics[width=\textwidth]{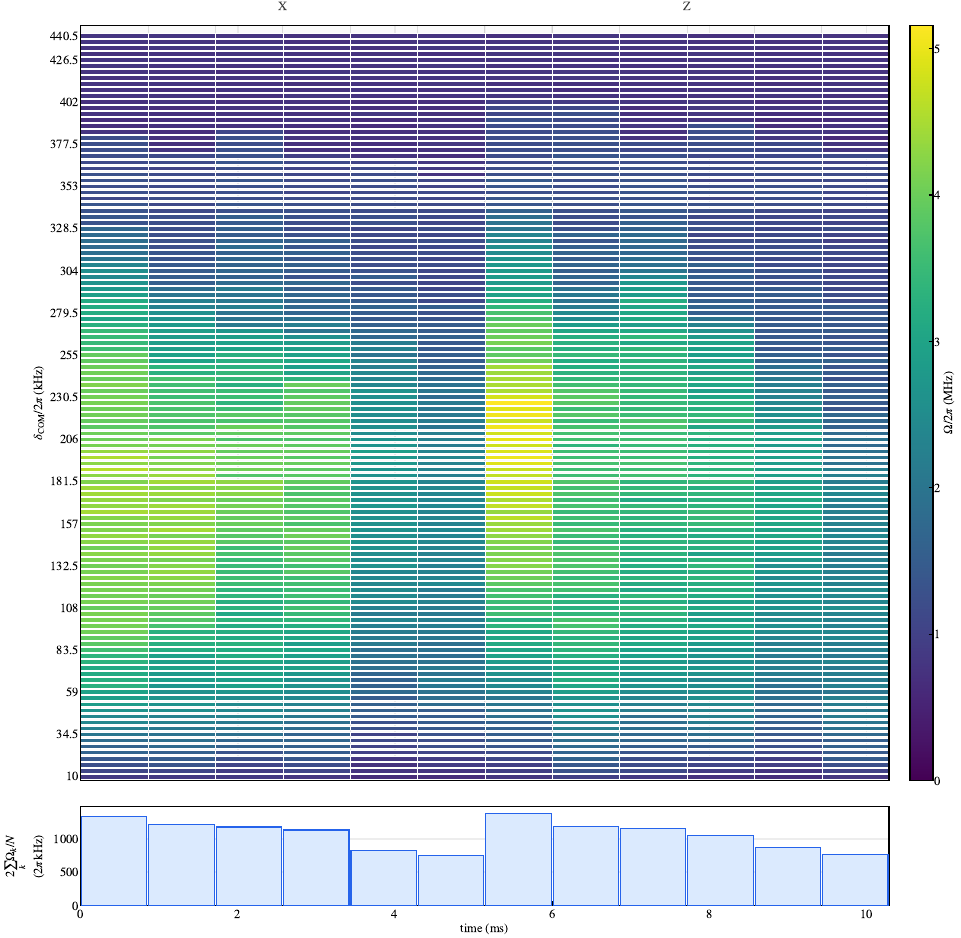}
\caption{\label{fig:raman_ms_schedule_012}
Representative schedule with the minimum \(S=12\) time slots and \(T_{\rm cycle}=10.3\,\mathrm{ms}\).  The peak per-ion effective-Rabi allocation is \(2\sum_k\Omega_k/N=2\pi\times1378.7\,\mathrm{kHz}\).}
\end{figure}

Figures~\ref{fig:raman_ms_schedule_104}, \ref{fig:raman_ms_schedule_036}, and \ref{fig:raman_ms_schedule_012} visualize three settings of this tradeoff.  The upper panel of each figure displays every scheduled gate as a colored horizontal bar, with its row giving the frequency-band offset \(\delta_{\rm COM}/2\pi\).  The lower panel reports the effective-Rabi allocation per ion, \(2\sum_k\Omega_k/N\) for each time slot.  The three schedules span the long-duration low-power regime, an intermediate schedule, and the shortest possible schedule, while retaining the common gate duration \(T=857.14\,\mu\mathrm{s}\).

\clearpage

\section{Circuit-level simulation and decoding of a BB18 memory}
\subsection{Code and memory circuit}
We use the bivariate-bicycle (BB) CSS-code construction~\cite{bravyi2024_high_threshold_bb_memory}.  Let \(A=a(x,y)\) and \(B=b(x,y)\) denote the \(lm\times lm\) binary circulant matrices obtained from two polynomials in \(\mathbb F_2[x,y]/(x^l+1,y^m+1)\).  The parity-check matrices are
\begin{equation}
H_X=\begin{bmatrix}A&B\end{bmatrix},
\qquad
H_Z=\begin{bmatrix}B^{\mathsf T}&A^{\mathsf T}\end{bmatrix}.
\label{eq:bb_parity_checks}
\end{equation}
For the BB18 code, \(l=31\), \(m=4\), and the defining polynomials are~\cite{cain2026_shor_10000_reconfigurable_atoms}
\begin{equation}
a(x,y)=1+x^6y+x^{27},
\qquad
b(x,y)=y^2+x^{15}y^3+x^{24}.
\label{eq:bb248_polynomials}
\end{equation}
This gives the weight-six \([[248,10,18]]\) code, with 248 data qubits and 124 checks of each type.

We simulate a 9-round \(X\)-memory experiment.  The data qubits are initialized in \(\lvert+\rangle^{\otimes248}\).  An \(X\)-check ancilla is prepared in \(\lvert+\rangle\), acts as the control of its six CNOTs, and is measured in the \(X\) basis; a \(Z\)-check ancilla is prepared in \(\lvert0\rangle\), acts as the target of its six CNOTs, and is measured in the \(Z\) basis.  Detection events are formed by comparing consecutive syndrome outcomes, together with the initial and final time boundaries.  After the ninth round, the data qubits are measured in the \(X\) basis and their parities define the ten logical-\(X\) observables.

The \(X\)-memory is the conservative benchmark for the noise model considered here.  Idle dephasing contributes only \(Z\) faults and is one of the dominant error mechanisms, while the remaining gate error and crosstalk channels are balanced between \(X\)- and \(Z\)-type endpoints when aggregated over the syndrome circuit.  The \(X\)-memory is directly sensitive to these additional idle-\(Z\) faults.  Thus, within this model and in the absence of another basis-dependent mechanism, the corresponding \(Z\)-memory is expected to perform no worse.

\subsection{Pauli-twirled physical noise model}
The Clifford simulation samples five classes of physical errors: idle spin dephasing, two-qubit-phase error, crosstalk between parallel gates, residual spin--phonon entanglement, and measurement errors.  We first convert each physical error into a stochastic Pauli channel.  For an idle interval of duration \(t\), dephasing with coherence time \(T_2\) gives
\begin{equation}
\mathcal E_{\rm idle}(\rho)=(1-p_Z)\rho+p_Z Z\rho Z,
\qquad
p_Z(t,T_2)=\frac{1-e^{-t/T_2}}{2}.
\label{eq:idle_pauli_channel}
\end{equation}

If the implemented entangling angle has a $\Delta\theta$ deviation from the ideal angle $\frac{\pi}{4}$, it will result in the following two-qubit-phase error after Pauli-twirling:
\begin{equation}
\mathcal E_{\rm tar}(\rho)
=\left(1-p_{\theta}\right)\rho
+p_{\theta}(X_cX_t)\rho(X_cX_t),
\qquad
p_{\theta}=\sin^2\!\Delta\theta.
\label{eq:target_miss_pauli_channel}
\end{equation}
The local Clifford rotations that realize a CNOT${}_{\rm ct}$ map the native two-qubit error to \(Z_cX_t\).  Likewise, a non-target phase \(\theta_{ij}^{\rm xtalk}\) accumulated between qubits \(i\) and \(j\) of two simultaneous gates gives
\begin{equation}
p_{ij}^{\rm xtalk}=\sin^2\!\theta_{ij}^{\rm xtalk}.
\label{eq:xtalk_pauli_probability}
\end{equation}
Each endpoint is mapped according to its role in its own CNOT: the native \(X\)-axis operator becomes \(Z\) on a control and \(X\) on a target.  A crosstalk channel can therefore produce \(ZZ\), \(ZX\), \(XZ\), or \(XX\) in the CNOT frame~\cite{liu2025_parallel_gate_crosstalk}.

For residual spin--phonon entanglement, let \(\alpha_{im}\) be the residual displacement of mode \(m\) conditioned on qubit \(i\), and let \(\bar n_m\) be the thermal occupation of that mode.  The residual evolution generated by the first-order Magnus term at the end of a gate is
\begin{equation}
U_{\rm disp}
=
\exp\left[
 \sum_{i,m}X_i
\left(\alpha_{im}a_m^\dagger-\alpha_{im}^*a_m\right)
\right].
\label{eq:residual_displacement_unitary}
\end{equation}
The two eigenstates of \(X_i\) therefore condition mode \(m\) on displacements that differ by \(2\alpha_{im}\).  With \(D(\gamma)=\exp(\gamma a_m^\dagger-\gamma^*a_m)\), a thermal mode satisfies
\begin{equation}
\operatorname{Tr}\!\left[D(\Delta\alpha)\rho_{{\rm th},m}\right]
=
\exp\left[-\frac{2\bar n_m+1}{2}
\lvert\Delta\alpha\rvert^2\right].
\label{eq:thermal_displacement_overlap}
\end{equation}
More generally, let \(X_i\lvert\boldsymbol{s}\rangle=s_i\lvert\boldsymbol{s}\rangle\), with \(s_i=\pm1\).  After tracing out all modes, the magnitude of an \(X\)-basis coherence is attenuated by
\begin{equation}
\left|
\frac{\rho_{\boldsymbol{s},\boldsymbol{s}'}^{\rm out}}
{\rho_{\boldsymbol{s},\boldsymbol{s}'}^{\rm in}}
\right|
=
\exp\left[
-\frac{1}{2}\sum_m(2\bar n_m+1)
\left|
\sum_i(s_i-s_i')\alpha_{im}
\right|^2
\right].
\label{eq:residual_multiqubit_dephasing}
\end{equation}
For a coherence whose two spin configurations differ only at qubit \(i\), this reduces to \(\exp(-2C_{ii})\), where
\begin{equation}
C_{ii}=\sum_m(2\bar n_m+1)\lvert\alpha_{im}\rvert^2,
\label{eq:residual_dephasing_exponent}
\end{equation}
whereas a stochastic \(X\)-axis Pauli channel
\begin{equation}
\mathcal E_i^{\rm res}(\rho)
=
(1-p_i^{\rm res})\rho
+p_i^{\rm res}X_i\rho X_i
\end{equation}
attenuates the same coherence by \(1-2p_i^{\rm res}\).  Equating these two attenuation factors gives
\begin{equation}
p_i^{\rm res}=\frac{1-e^{-2C_{ii}}}{2}
\approx C_{ii},
\label{eq:residual_pauli_probability}
\end{equation}
where the last expression applies when \(C_{ii}\ll1\).  Finally, the CNOT-frame Clifford rotations map the native \(X_c\) fault to \(Z_c\) and \(X_t\) to \(X_t\).  Equations~(\ref{eq:idle_pauli_channel})--(\ref{eq:residual_pauli_probability}) convert \(T_2\), \(\Delta\theta\), \(\theta_{ij}^{\rm xtalk}\), and \(\alpha_{im}\) into the Bernoulli probabilities sampled by the Clifford simulation.  The resulting Pauli frames are then propagated exactly through the syndrome circuit.

In the main text, the two noise settings use \(\bar n_m=1\) and \(0.1\), respectively, for every transverse mode.  We apply the same static offset \(\Delta\omega\) to all mode frequencies while holding the beat notes and zero-drift gate calibrations fixed.  The gate duration, beat-note assignment and calibrated \(\Omega_{e,k}\) are fixed by the schedule. They determine \(\Delta\theta\), \(\theta_{ij}^{\rm xtalk}\), and \(\alpha_{im}\) channel by channel.  Independent bit flips are applied immediately before each ancilla measurement and the final data-qubit readout, with \(p_{\mathrm{meas}}=10^{-3}\) and \(5\times10^{-4}\) for the two parameter sets, respectively.  We neglect preparation faults and single-qubit gate errors.

\subsection{Circuit-fault coordinates and DEM compression}
In circuit-level noise coordinates, a binary variable \(u_j\) labels a particular primitive fault mechanism at a particular circuit location.  This representation retains its source-specific probability, but it becomes very large under the present crosstalk model.  A slot containing \(m\) parallel two-qubit gates has \(2m\) endpoints and
\begin{equation}
\binom{2m}{2}-m=2m(m-1)
\label{eq:number_xtalk_channels}
\end{equation}
distinct cross-gate qubit pairs.  The number of circuit-fault variables therefore grows quadratically with the degree of parallelism.

To construct the detector and logical effects of these faults, we use the phenomenological noise model
\begin{equation}
x_{\rm prop}=(e_1,\ldots,e_{R+1},m_1,\ldots,m_R),
\label{eq:propagated_coordinate_vector}
\end{equation}
where \(e_t\in\mathbb F_2^n\) is the data-\(Z\) error added at the \(t\)-th cycle, and \(m_t\in\mathbb F_2^{r_X}\) is the \(X\)-check measurement error of the same round.  Setting \(m_0=m_{R+1}=0\), the detector vector \(d=(d_1,\ldots,d_{R+1})\in\mathbb F_2^{(R+1)r_X}\) records changes between consecutive \(X\)-check outcomes, with
\begin{equation}
d_t=H_Xe_t+m_t+m_{t-1}.
\label{eq:propagated_detector_relation}
\end{equation}
The logical-action vector \(\ell\in\mathbb F_2^{k_X}\) records which logical \(X\) observables are flipped by the accumulated data-\(Z\) error.  If \(L_X\in\mathbb F_2^{k_X\times n}\) maps a data-\(Z\) error to its logical action, then
\begin{equation}
d=H_{\rm prop}x_{\rm prop},
\qquad
\ell=A_{\rm prop}x_{\rm prop},
\label{eq:propagated_detector_logical_action}
\end{equation}
where
\begin{equation}
H_{\rm prop}
=\begin{bmatrix}
I_{R+1}\otimes H_X & D_R\otimes I_{r_X}
\end{bmatrix},
\qquad
A_{\rm prop}
=\begin{bmatrix}
\mathbf 1_{R+1}^{\mathsf T}\otimes L_X & 0
\end{bmatrix}.
\label{eq:propagated_detector_logical_matrices}
\end{equation}
Here \(D_R\in\mathbb F_2^{(R+1)\times R}\) is the incidence matrix of the time path, so its column \(t\) has ones in detector layers \(t\) and \(t+1\).  Clifford propagation supplies the linear map \(x_{\rm prop}=M_{u\to x}u\), and hence
\begin{equation}
H_{\rm raw}=H_{\rm prop}M_{u\to x},
\qquad
A_{\rm raw}=A_{\rm prop}M_{u\to x},
\label{eq:raw_detector_logical_matrices}
\end{equation}
so that a sampled circuit-level error vector produces \(d=H_{\rm raw}u\) and \(\ell=A_{\rm raw}u\).  This construction is used only to export the detector and logical effects entering the DEM.  For example, the unpruned 9-round \(S=36\) circuit-fault cache contains \(1{,}143{,}867\) variables, which reduce to \(186{,}029\) distinct nontrivial detector--logical effects after removing faults that affect neither a detector nor a logical observable and merging faults with identical effects.

For the reported DEM-\(u\) results, if a merged class \(g\) contains independent primitive faults with probabilities \(p_j\), its exact odd-parity probability is
\begin{equation}
q_g=\frac{1-\prod_{j\in g}(1-2p_j)}{2}.
\label{eq:merged_dem_probability}
\end{equation}
Our simulation and decoding procedure is therefore:
\begin{enumerate}
\item Use Clifford propagation to construct \(H_{\rm raw}\) and \(A_{\rm raw}\), then construct the compressed DEM-\(u\) groups and their priors \(q_g\).
\item Sample each physical shot in circuit-fault coordinates from the complete five-source Bernoulli model, and obtain its detector and logical ground truth as \(d=H_{\rm raw}u\) and \(\ell=A_{\rm raw}u\).
\item Decode \(d\) in the compressed DEM-\(u\) coordinates to obtain an estimated merged-fault vector \(\hat u_{\rm DEM}\).
\item Declare success when the decoded and true logical actions agree, \(A_{\rm DEM}\hat u_{\rm DEM}=A_{\rm raw}u\).
\end{enumerate}
Thus physical faults are sampled with their circuit-level probabilities, while decoding retains their exact combined priors without keeping duplicate detector--logical columns.

\subsection{RelayBP parameters}
We decode the compressed DEM-\(u\) graph using RelayBP~\cite{muller2025_relay_bp_realtime_memory} and its open-source implementation~\cite{alexander2026_relay_bp_software}.  RelayBP augments MinSum belief propagation with a memory-dependent variable-node bias,
\begin{equation}
\Lambda_i^{(k)}=(1-\gamma_i)L_i+\gamma_iM_i^{(k-1)},
\label{eq:relaybp_memory_bias}
\end{equation}
where \(L_i\) is the prior log likelihood and \(M_i^{(k-1)}\) is the previous marginal.  Different memory strengths \(\gamma_i\) break local symmetries and help the decoder escape trapping sets; successive relay legs inherit the preceding leg's final marginals.

All reported simulations use the single-precision \texttt{RelayDecoderF32} implementation with the merged physical priors in Eq.~(\ref{eq:merged_dem_probability}).  The first leg uses \(\gamma_0=0.1\) for at most 80 iterations.  Each later leg independently draws every \(\gamma_i\) from the uniform distribution \(\mathcal U[-0.24,0.66]\); we allow at most 300 such legs and 60 iterations per leg.  Decoding stops after the first syndrome-consistent candidate is found (RelayBP-1).  The optional MinSum message-scaling coefficient is left at the library default (\texttt{alpha=None}), its iteration scaling factor is 1, and the relay random seed is 0.

\subsection{Logical-error-rate extrapolation}
At the target physical parameters, direct Monte Carlo sampling becomes impractical when logical failures are rare.  Following established low-error-rate extrapolation procedures~\cite{bonillaataides2021_xzzx_surface_code,pecorari2025_high_rate_ldpc_neutral_atoms,cain2026_shor_10000_reconfigurable_atoms}, we multiply every circuit-fault probability by a common factor \(\lambda\),
\begin{equation}
p_j(\lambda)=\lambda p_j,
\label{eq:uniform_noise_scaling}
\end{equation}
using scale factors for which all probabilities remain below one.  For each \(\lambda>1\), we independently sample the full five-source noise model, decode in compressed DEM-\(u\) coordinates, and measure the 9-cycle block logical error probability \(p_L^{(9)}(\lambda)\).  At each \(\lambda\), sampling continues until \(k=50\) logical failures have been observed.  If \(T_k\) denotes the total number of samples required to observe the \(k\)th failure, we estimate the failure probability using the unbiased negative-binomial estimator
\begin{equation}
\widehat p_L^{(9)}(\lambda)=\frac{k-1}{T_k-1}.
\label{eq:negative_binomial_logical_error_estimator}
\end{equation}
The sampled points are fitted to
\begin{equation}
\log p_L^{(9)}(\lambda)=a+b\log\lambda,
\qquad
p_L^{(9)}(1)=e^a.
\label{eq:logical_error_extrapolation}
\end{equation}
The reported per-cycle logical error rate is converted from the block logical error rate using
\begin{equation}
p_L(1)=1-\left[1-p_L^{(9)}(1)\right]^{1/9}.
\label{eq:block_to_cycle_logical_error}
\end{equation}

\begin{figure}[t]
\centering
\includegraphics{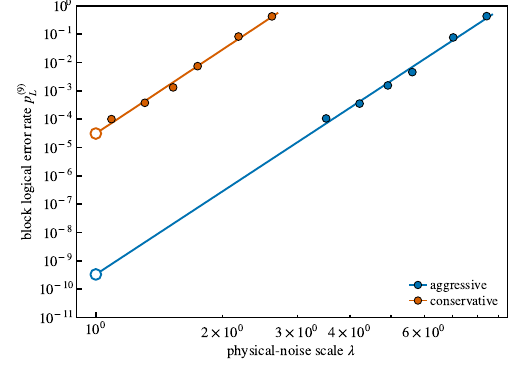}
\caption{\label{fig:slots036_dem_extrapolation}
DEM-\(u\) logical-error-rate extrapolation for the 9-cycle BB18 \(X\)-memory experiment with the \(S=36\) schedule.  Blue and orange denote the aggressive and conservative physical parameter sets, respectively.  Filled circles are direct Monte Carlo estimates, solid lines are the log--log fits in Eq.~(\ref{eq:logical_error_extrapolation}), and open circles show the extrapolated values at \(\lambda=1\).}
\end{figure}

Figure~\ref{fig:slots036_dem_extrapolation} shows the DEM-\(u\) extrapolation for the representative \(S=36\) schedule.  The fitted slopes \(b\) are \(9.77\) and \(9.87\) for the aggressive and conservative parameter sets, respectively.  The corresponding extrapolated per-cycle logical error rates at \(\lambda=1\) are \(3.7\times10^{-11}\) and \(3.4\times10^{-6}\).

